\RequirePackage{fix-cm}

\documentclass[onecolumn]{svjour3} 
\smartqed  
\usepackage{graphicx}
\usepackage{amssymb} 

\usepackage{subfigure}
\usepackage{amsmath}
\usepackage{amsfonts}
\usepackage{amsbsy}
\usepackage{times}
\usepackage[left]{lineno}

 \journalname{J. Braz. Soc. Mech. Sci. Eng.}
\begin{document}

\title{Length scale of density waves in the gravitational flow of fine grains in pipes\thanks{Accepted Manuscript for the Journal of the Brazilian Society of Mechanical Sciences and Engineering, v. 37, p. 1507-1513, 2015. The final publication is available at Springer via http://dx.doi.org/10.1007/s40430-014-0291-3}}


\author{Erick de Moraes Franklin \and
        Carlos Alvarez Zambrano 
}


\institute{Erick de Moraes Franklin \at
              Faculty of Mechanical Engineering, University of Campinas - UNICAMP \\
              Tel.: +55-19-35213375\\
              \email{franklin@fem.unicamp.br}            \\    
									 \and
           Carlos Alvarez Zambrano \at
              Faculty of Mechanical Engineering, University of Campinas - UNICAMP \\
							 \email{calvarez@fem.unicamp.br}      
}

\date{Received: date / Accepted: date}

\maketitle

\begin{abstract}
Gravitational flow of grains in pipes is frequently encountered in industry. When the grains and pipes are size-constrained, granular flow may result in density waves consisting of alternate high- and low-compactness regions. This paper discusses the length scale of density waves that appear when fine grains fall vertically in pipes. A one-dimensional model and a linear stability analysis of the model are presented. The analysis suggests the presence of long-wavelength instability for the most unstable mode, moreover, a cut-off wavenumber from which the length scale is estimated. Finally, the model results are compared to experimental data.
\end{abstract}

\section{List of Symbols}
\label{section:nomenclature}

$A_1$ to $A_9$ = constants;\\
$a$ = constant;\\
$B$ = constant;\\
$B_1$ to $B_5$ = constants;\\
$b$ = constant;\\
$C_1$ to $C_5$ = constants; \\
$c$ = granular compactness; \\
$D$ = tube diameter ($m$); \\
$d$ = grain diameter ($m$); \\
$g$ = gravity acceleration ($m/s^2$); \\
$H$ = humidity index; \\
$k$ = wavenumber ($m^{-1}$); \\
$P$ = pressure ($Pa$);\\
$P_{atm}$ = atmospheric pressure ($Pa$);\\
$R$ = tube radius ($m$); \\
$v_s$ = velocity of individual grains ($m/s$);\\
$W$ = grain flow rate ($kg/s$);\\
$z$ = vertical coordinate ($m$). \\

\noindent \textbf{Greek symbols}\\
$\kappa$ = redirection coefficient; \\
$\gamma$ = ratio of specific heats; \\
$\lambda$ = wavelength of the plugs ($m$);\\
$\mu_a$ = dynamic viscosity of air ($Pa.s$); \\
$\mu_s$ = friction coefficient between grains; \\
$\rho_{s}$ = specific mass of each grain ($kg/m^3$);\\
$\omega_r$ = angular frequency ($rad/s$);\\
$\omega_i$ = growth rate ($s^{-1}$);\\
$\sigma_{zr}$ = stress between the tube wall and the grains($N/m^2$);\\
$\sigma_{zz}$ = vertical stress operating on the grains ($N/m^2$).\\

\noindent \textbf{Subscripts}\\
$a$ = relative to air;\\
$s$ = relative to grains;\\
$0$ = relative to the basic state.\\

\noindent \textbf{Superscripts}\\
$\tilde{}$ = relative to the perturbation;\\
$\hat{}$ = relative to the amplitude of perturbations.\\

\section{Introduction}

Granular matter is abundant on Earth, e.g., 20\% of the Earth surface consists of sand and other solid fragments \cite{Duran}. As a result, gravitational flow of these materials is frequently observed in nature and in industry. Nonetheless, the behavior of granular flows is not well understood and the rheology of granular media is a matter of debate. Given the importance of granular flow in nature and industry, considerable work has been done to understand the dynamics and instabilities of granular media \cite{Campbell,Elbelrhiti,Franklin_6,GDR_midi,Jaeger}.

In industry, granular gravitational flows generally occur in pipes or closed conduits, e.g., the transport of grains to silos in the food industry, the transport of sand in civil constructions, and the transport of powders in the chemical industry. When the grains and the tube diameter are size-constrained, granular flow may give rise to instabilities. These commonly undesired instabilities consist of alternate regions of high and low compactness (grain concentration), and are characterized by intermittency, oscillating patterns and blockages \cite{Aider,Bertho_1,Raafat}. 

This kind of instability may appear in vacuum, or when the effect of air is negligible \cite{Savage,Wang}. However, in the case of fine grains, these patterns are recognized as products of the interaction between falling grains and trapped air. Raafat et al. (1996) \cite{Raafat} studied the formation of density waves in pipes experimentally. The experiments were performed in a $1.3\,m$ long tube with an internal diameter $D$ of $2.9\,mm$ using glass splinters and glass beads with mean grain diameter $d$ of $0.09\,mm$ to $0.2\,mm$ and $0.2\,mm$, respectively. They observed density waves for moderate grain flow rate and when the ratio between the pipe and the grain diameter is $6 \leq D/d \leq 30$. Furthermore, they proposed that the friction between the grains and the forces between the trapped air and the grains are responsible for the density waves.

Aider et al. (1999) \cite{Aider} presented an experimental study of the granular flow patterns in vertical pipes. The experiments were performed in a tube similar with that of Raafat et al. (1996) \cite{Raafat} using glass beads with mean diameter of $0.125\,mm$. The density variations were measured by using a linear CCD (charge coupled device) camera and a frequency of up to $2\,kHz$. Aider et al. (1999) \cite{Aider} observed that the density waves consisted of high-compactness plugs ($c\approx 60\%$, where $c$ is the compactness) separated by low-density regions; furthermore, the density waves appeared when the grain flow rate $W$ was $1.5\,g/s\,-\, 2.5\,g/s$ (oscillating waves) or $2.5\,g/s\,-\,5\,g/s$ (propagative waves). These authors also noted that humidity $H$ must be within $35\%$ and $75\%$, otherwise the grains clogged the tube owing to capillary forces ($H>75\%$) or due to electrostatic forces ($H<35\%$).

Bertho et al.(2002) \cite{Bertho_1} presented experiments on density waves performed using an experimental setup similar to that of Raafat et al. (1996) \cite{Raafat} and Aider et al. (1999) \cite{Aider}. The vertical tube ($D=3mm$, $1.25m$ long) and the glass beads ($d\,=\,0.125\,mm$ glass beads) were roughly the same as that of Aider et al. (1999) \cite{Aider}, and a linear CCD camera was used. In addition, capacitance sensors were used to measure the compactness of grains at two different locations, and the pressure distribution was also measured. The experimental data showed that the characteristic length of the high-compactness regions of the density wave regime is in the order of $10\,mm$.

Ellingsen et al. (2010) \cite{Ellingsen} studied the gravitational flow of grains in a narrow pipe under vacuum conditions. They performed numerical simulations based on a one-dimensional model for the granular flow where the collisions were modeled by two coefficients of restitution, one among grains and the other between the grains and the pipe walls. A narrow pipe was assumed and periodic boundary conditions were employed. The numerical results showed that granular waves could form in the absence of air if the dissipation caused by the collisions among the grains were smaller than those between the grains and the walls. However, the proposed model cannot predict the wavelength of the density waves in the presence of interstitial gas.

This paper discusses the length scale of density waves that appear when fine grains fall through a vertical pipe in the presence of air. It presents a one-dimensional flow model based on the work of Bertho et al. (2003) \cite{Bertho_2} and a linear stability analysis of the flow. The model results are then compared to experimental data. To the best of our knowledge, this is the first time that a simple stability analysis allows the prediction of the correct length scale in such a problem.

The next section describes the physics and the main equations of the one-dimensional model. The following sections present the initial stability analysis of the granular flow, the performed experiments, and the main results, respectively. The conclusion section follows.

\section{One-dimensional two-phase model}
\label{section:model}

\subsection{Granular flow in a vertical pipe}

The problem analyzed here consists of cohesionless grains falling from a hopper in a tube. The ratio between the mean grain diameter and the tube diameter is within $6 \leq D/d \leq 30$ and the humidity is between $35<H<75\%$. The grain size and specific mass are such that the air effects are not negligible, a typical case is the $d\,=\,0.125\,mm$ glass beads, as in Aider et al. (1999) \cite{Aider} and Bertho et al. (2002) \cite{Bertho_1}. In this case, density waves are expected for moderate grain flow rates.

\begin{figure}
  \begin{center}
    \includegraphics[width=.85\linewidth]{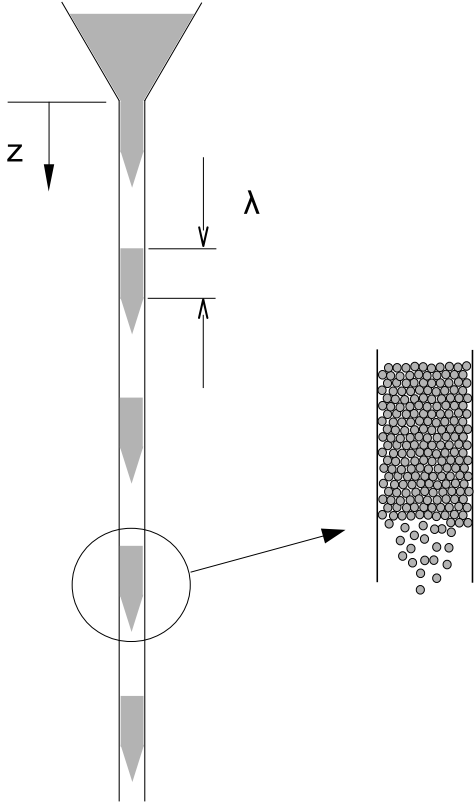}
    \caption{Layout of the gravitational granular flow in a tube. The high and low grain concentration regions are shown. $z$ is the vertical coordinate and $\lambda$ is the wavelength of the granular plugs.}
    \label{fig:escopo}
  \end{center}
\end{figure}

The density waves consist of alternate regions of high grain concentration, which are essentially plugs of granular material, and low grain concentration, which are air bubbles with dispersed free-falling grains. In the high-concentration regions, the compactness is assumed close to its maximum value, $c\approx 60\%$; therefore, grains in the plug periphery are in contact with the tube wall. Under these conditions, the redirection of forces is present within the plug and the Janssen effect \cite{Duran} is expected if the plugs are long enough. In the low-concentration regions, the air pressure increases owing to the stresses caused by the neighboring plugs as well as the volume decrease (compression) caused by the free-falling grains. Figure \ref{fig:escopo} shows the layout of the gravitational granular flow.

\subsection{One-dimensional model}

To analyze the problem, a one-dimensional model based on the work of Bertho et al. (2003) \cite{Bertho_2} was used. The modifications proposed in this study concern the closure equations for the stresses within the grains and the inclusion of a parameter that considers the capillary forces that were not considered in Bertho et al. (2003) \cite{Bertho_2}.

The main objective of the analysis in Section \ref{section:analysis} is to find the characteristic length of the plugs. Therefore, we apply the one-dimensional model to the plug region. The model consists of an equation for the pressure of air, which flows through the plug from one neighboring bubble to the next, and of a motion equation for the grains within the plug.

\subsubsection{Air pressure}

Bertho et al. (2003) \cite{Bertho_2} combined the mass conservation equations for the air and the grains, the isentropic relation for the air, and Darcy's equation relating the air flow through packed grains to the pressure gradient to obtain Eq. \ref{eq:pressure}

\begin{equation}
\frac{\partial P}{\partial t} + v_s\frac{\partial P}{\partial z} + \frac{\gamma P}{(1-c)}\frac{\partial v_s}{\partial z} - B\frac{\partial^2 P}{\partial z^2}=0
\label{eq:pressure}
\end{equation}

\noindent where $P$ is the pressure, $v_s$ is the velocity of the individual grains, $c$ is the granular compactness, $z$ is the vertical coordinate, $\gamma$ is the ratio of specific heats ($1.4$ for air), and $B$ is a coefficient given by

\begin{equation}
B = \frac{\gamma P\left( 1-c\right)^2d^2}{\mu_a180c^2}
\label{eq:pressure_2}
\end{equation}

\noindent where $\mu_a$ is the dynamic viscosity of air.

\subsubsection{Grain motion}

The equation of motion for the grains is given by

\begin{equation}
\rho_sc\left( \frac{\partial v_s}{\partial t} + v_s\frac{\partial v_s}{\partial z} \right) = \rho_scg - \frac{\partial P}{\partial z}  - \frac{\partial \sigma_{zz}}{\partial z} - \frac{2}{R}\sigma_{zr}
\label{eq:motion}
\end{equation}

\noindent where $\rho_s$ is the specific mass of each grain, $g$ is the gravitational acceleration, $R$ is the tube radius and $\sigma_{mn}$ is the stress at the surface $n$ in the $m$ direction. In this manner, $\sigma_{zz}$ is the vertical stress operating on the grains and $\sigma_{zr}$ is the stress between the tube wall and the grains.

Two modifications are proposed for the closure of $\sigma_{zz}$ and $\sigma_{zr}$ in the present model. The first is to model $\sigma_{zr}$ as a function of the square of the grain velocity

\begin{equation}
\sigma_{zr} \sim \rho_s\mu_s v_s^2
\label{eq:proposition}
\end{equation}

\noindent where $\mu_s\approx\tan(32^o)$ is the friction coefficient between grains. The redirection of forces needs also to be considered. Typically, \cite{Duran}, this is done through a constant coefficient $\kappa$: $\sigma_{zr}=\mu_s\kappa\sigma_{zz}$. Saturation in stresses is attributed to the Janssen effect \cite{Duran} and taken into account in this study via an exponential function of the plug length. Similarly, capillary forces operate as bonding forces that are proportional to the plug length and are modeled as an exponential function of the plug length as well. This is the second modification. Finally, we obtain Eqs. \ref{eq:closure_1} and \ref{eq:closure_2} for the closure.

\begin{equation}
\sigma_{zr} \sim \frac{1}{2}\rho_s\mu_s v_s^2 b\exp (-ak )
\label{eq:closure_1}
\end{equation}

\begin{equation}
\frac{\partial \sigma_{zz}}{\partial z}  \sim \frac{\rho_s}{\kappa} v_s \frac{\partial v_s}{\partial z} b\exp (-ak )
\label{eq:closure_2}
\end{equation}

\noindent where $a$ and $b$ are constants, $k=2\pi\lambda^{-1}$ is the wavenumber, and $\lambda$ is the wavelength of the plugs. Constant $b$ is directly proportional to the air water surface tension and models the capillary force.

\section{Stability analysis}
\label{section:analysis}

The stability analysis was performed based on Eqs. \ref{eq:pressure} and \ref{eq:motion} with the closure Eqs. \ref{eq:closure_1} and \ref{eq:closure_2}. The main objective was to find the wavelength for the high-density plugs of the granular flow in the pipe. As the plugs have a constant compactness $c$ of $\approx 60\%$, $c$ is considered constant. Equations \ref{eq:pressure} and \ref{eq:motion} are then solved for $P$ and $v_s$.

The analysis considers a basic state in which the pressure is equal to the atmospheric pressure, $P_0 = P_{atm}$, and the grain velocity is equal to the mean velocity obtained from the mass flow rate, $v_0=4W(c\rho_s\pi D^2)^{-1}$. The pressure and grain velocity are then the sum of the basic state and the perturbation, the latter is assumed much smaller than the corresponding basic state,

\begin{equation}
P=P_0 + \tilde{P},\ \ v_s=v_0 + \tilde{v}
\label{eq:basic_pert}
\end{equation}

\noindent where $\tilde{P}$ and $\tilde{v}$ are respectively the pressure and velocity perturbations. $P_0/P_{atm}$ and $v_0/(4w(c\rho_s\pi D^2)^{-1})$ are $O(1)$ while $\tilde{P}/P_{atm}$ and $\tilde{v}/(4W(c\rho_s\pi D^2)^{-1})$ are $O(\epsilon)$, $\epsilon\ll 1$.

The linear analysis was performed by inserting the pressure and the velocity from Eq. \ref{eq:basic_pert} in Eqs. \ref{eq:pressure} and \ref{eq:motion}, and keeping only the terms of $O(\epsilon)$. The equations for $O(\epsilon)$ are then

\begin{equation}
\frac{\partial \tilde{P}}{\partial t} + v_0\frac{\partial \tilde{P}}{\partial z} + \frac{\gamma P_0}{(1-c)}\frac{\partial \tilde{v}}{\partial z} - B_1\frac{\partial^2 \tilde{P}}{\partial z^2}=0
\label{eq:pert_pressure}
\end{equation}

\begin{equation}
\rho_sc\left( \frac{\partial \tilde{v}}{\partial t} + v_0\frac{\partial \tilde{v}}{\partial z} \right) = - \frac{\partial \tilde{P}}{\partial z}  - B_3v_0\frac{\partial \tilde{v}}{\partial z} - B_5v_0\tilde{v}
\label{eq:pert_motion}
\end{equation}

\noindent In Eq. \ref{eq:pert_pressure}, $B_1$ is a constant given by

\begin{equation}
B_1 = \frac{\gamma P_0\left( 1-c\right)^2d^2}{\mu_a180c^2}
\label{eq:B1}
\end{equation}

\noindent In Eq. \ref{eq:pert_motion}, $B_3$ and $B_5$ are exponentially decaying functions of $k$ given by

\begin{equation}
B_3 =  \frac{\rho_s}{\kappa} b\exp (-ak )
\label{eq:B3}
\end{equation}

\begin{equation}
B_5 =  \frac{2\rho_s\mu_s}{D} b\exp (-ak )
\label{eq:B5}
\end{equation}

\noindent and are approximated as constants in the first part of the solution, which means that we are considering long waves. This is justified \textit{a posteriori}.

Equations \ref{eq:pert_pressure} and \ref{eq:pert_motion} form a linear system with solutions consisting of plane waves. The solutions can be found by considering the normal modes

\begin{equation}
\begin{array}{c} \tilde{P}\,=\,\hat{P}e^{i\left( kz-\omega t\right) }\,+ c.c. \\ \, \\ \tilde{v}\,=\,\hat{v}e^{i\left( kz-\omega t\right) }\,+ c.c. \\ \end{array}
\label{normal_modes}
\end{equation}

\noindent where $k\in \mathbb{R}$ is the wavenumber in the $z$ direction, $\hat{P}\in \mathbb{C}$ and $\hat{v}\in \mathbb{C}$ are the amplitudes, and $c.c.$ stands for the complex conjugate. Let $\omega\in \mathbb{C}$, $\omega\,=\,\omega_r\,+\,i\omega_i$, where $\omega_r\in \mathbb{R}$ is the angular frequency and $\omega_i\in \mathbb{R}$ is the growth rate. By inserting the normal modes in Eqs. \ref{eq:pert_pressure} and \ref{eq:pert_motion}, the following expression is obtained

\begin{equation}
\begin{array}{c} \left[\begin{array}{cc}-\omega+v_0ik+B_1k^2 & \frac{\gamma P}{(1-c)}ik \\ ik & \rho_sc\left( -i\omega +v_0ik\right) +B_3v_0ik+B_5v_0 \\ \end{array}\right] \left[\begin{array}{c}\hat{P} \\ \hat{v} \\ \end{array}\right] \,= \\ \, \\= \,\left[\begin{array}{c}0 \\ 0 \\ \end{array}\right] \end{array}
\label{eq:eigenvalue}
\end{equation}

The existence of non-trivial solutions for this system requires its determinant to be zero. This results in

\begin{equation}
\left\{ \begin{array}{c} -\omega_r^2 + \omega_i^2 + A_7k\omega_r + \left( (B_5v_0)/(\rho_sc)+B_1k^2\right)\omega_i + A_8k^2 = 0 \\ \, \\ \omega_r\left( A_2k^2+A_3+A_4\omega_i\right) - A_1k\omega_i - A_6k^3 - A_5k = 0 \\ \end{array}\right.
\label{eq:det}
\end{equation}

\noindent where

\begin{equation}
\left\{ \begin{array}{c} A_1 = B_3v_0 + 2\rho_scv_0 \\ \, \\ A_2 = \rho_scB_1 \\ \, \\ A_3 = B_5v_0 \\ \, \\ A_4 = 2\rho_sc \\ \, \\ A_5 = B_5v_0^2 \\ \, \\ A_6 = \rho_scv_0B_1 + B_3B_1v_0 \\ \, \\ A_7 = (B_3v_0)/(\rho_sc) + 2v_0 \\ \, \\ A_8 = -v_0^2 - (B_3v_0^2)/(\rho_sc) + v_0(B_5B_1)/(\rho_sc) + (\gamma P_0)/((1-c)\rho_sc)\\ \end{array}\right.
\label{eq:ctes}
\end{equation}

The system given by Eq. \ref{eq:det} can be solved  for $\omega_i$. The result is

\begin{equation}
C_1\omega_i^4 + C_2\omega_i^3 + C_3\omega_i^2 + C_4\omega_i + C_5 = 0
\label{eq:omega_i}
\end{equation}

\noindent where, for $A_9=(B_5v_0)/(\rho_sc)$,

\begin{equation}
\left\{ \begin{array}{c} C_1 = A_4^2 \\ \, \\ C_2 = (2A_2A_4+A_4^2B_1)k^2 + 2A_3A_4+A_4^2A_9 \\ \, \\ C_3 = (A_2^2 + 2A_2A_4B_1)k^4 + (-A_1^2+A_7A_1A_4 + 2A_3A_4B_1 + \\ + 2A_2A_4A_9 + A_4^2A_8)k^2 + 2A_2A_3+A_3^2+2A_3A_4A_9 \\ \, \\ C_4 = A_2^2B_1k^6 + (-2A_1A_6 + A_7A_6A_4 + 2A_2A_3B_1 + A_2^2A_9 + \\ + 2A_8A_2A_4)k^4 + A_7A_1A_2k^3 + (-2A_1A_5 + A_7A_1A_3 + A_7A_5A_4 + \\ + A_3^2B_1 + 2A_2A_3A_9 + 2A_8A_3A_4)k^2  + A_3^2A_9 \\ \, \\ C_5 = (-A_6^2+A_7A_6A_2+A_2^2A_8)k^6 + (-2A_5A_6+A_7A_5A_2 + \\ + A_7A_6A_3 +2A_2A_3A_8)k^4 + (-A_5^2+A_7A_5A_3+A_3^2A_8)k^2  \\ \end{array}\right.
\label{eq:ctes2}
\end{equation}

\section{Experiments}

\begin{figure}
  \begin{center}
    \includegraphics[width=.85\linewidth]{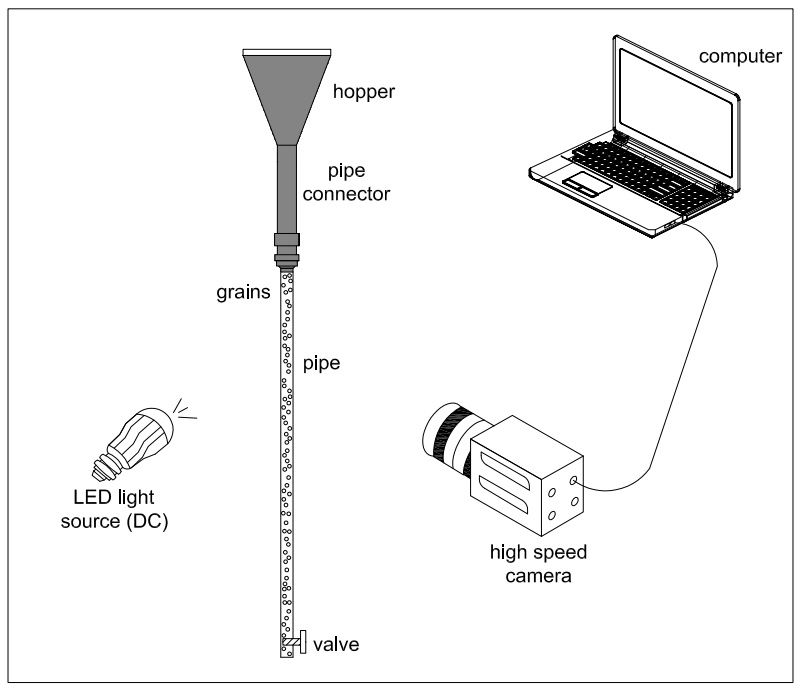}
    \caption{Experimental setup.}
    \label{fig:experimental_setup}
  \end{center}
\end{figure}

An experimental device was conceived and built to measure the length scale of density waves. The experimental device consisted of a storage reservoir, a hopper, a $1\,m$ long glass tube of $3\,mm$ internal diameter, and an exit valve. The tube was vertically aligned (within $\pm\,5^o$) and both the reservoir entrance and the exit valve were at atmosphere pressure. The grains consisted of glass beads of specific mass $\rho_s\,=\,2500\,kg/m^3$ divided in two different populations: grains with diameter within $212 \,\mu m\,\leq \, d \,\leq \, 300 \,\mu m$ and within $106 \,\mu m\,\leq \, d \,\leq \, 212 \,\mu m$. The temperature and the relative humidity were measured within $\pm\,0.5^oC$ and $\pm\,2.5\%$, respectively. The mass flow rate was measured by employing a chronometer and a $\pm\,0.01g$ accuracy balance. Figure \ref{fig:experimental_setup} presents a scheme of the experimental setup.

With the reservoir filled with grains, the exit valve was partially opened and the grains flowed in the tube. The migration of density waves was filmed with a $1280\,px \,\times\, 1024\,px$ high-speed camera (maximum frequency of $1000\,Hz$). In order to provide the necessary light for low exposure times while avoiding beating between the light source and the camera frequency, a grid of LED (Low Emission Diode) lamps was branched to a continuous current source. For the present experiments, the camera frequency was set to between $250\,Hz$ and $300\,Hz$. The number of acquired images for each test was $1500$ and the number of tests was $10$, giving a total of $15000$ images to be analyzed.

As soon as granular flow began, density waves were observed. The density waves had positive or zero mean celerity (the latter corresponding to purelly oscillating plugs), but the reasons for these different behaviors could not be identified in the present experiments. Aider et al. (1999) \cite{Aider} proposed that different celerity behaviors are due to different granular flow rates and humidity.

\begin{figure}
  \begin{center}
    \includegraphics[width=.55\linewidth]{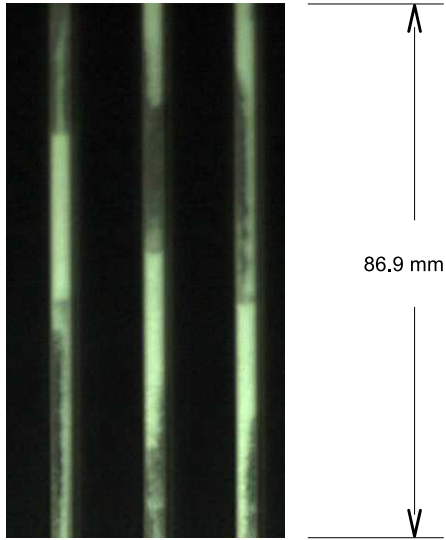}
    \caption{Density waves with positive celerity. The images were acquired at $250\,Hz$, but the time between frames in this figure is $0.06\,s$}
    \label{fig:exp1}
  \end{center}
\end{figure}

Figure \ref{fig:exp1} presents an example of density waves experimentally observed. The images, acquired at $250\,Hz$, show the presence of waves with positive celerity. In this figure, the time between frames is $0.06\,s$, and the wavelength is $\approx\,10D$.

\section{Results}
\label{section:Results}

Equation \ref{eq:omega_i} was numerically solved and only the results for which $\omega_i\in \mathbb{R}$ were considered pertinent. Constant $b$ in Eqs. \ref{eq:B3} and \ref{eq:B5} was considered $\sim \Gamma /D$, where $\Gamma$ is the surface tension of water. The rest of the model constants were fixed to the values used in Raafat et al. (1996) \cite{Raafat}, Aider et al. (1999) \cite{Aider}, and Bertho et al.(2002) \cite{Bertho_1}, i.e., $d=0.125mm$, $D=10mm$, $W=5g/s$, $\rho_s=2500 kg/m^3$, $c=0.6$, $\kappa =0.5$, $P_{atm}=10^5Pa$, and the gas properties were that of air.

\begin{figure}
   \begin{minipage}[c]{\textwidth}
    \begin{center}
      \includegraphics[scale=0.60]{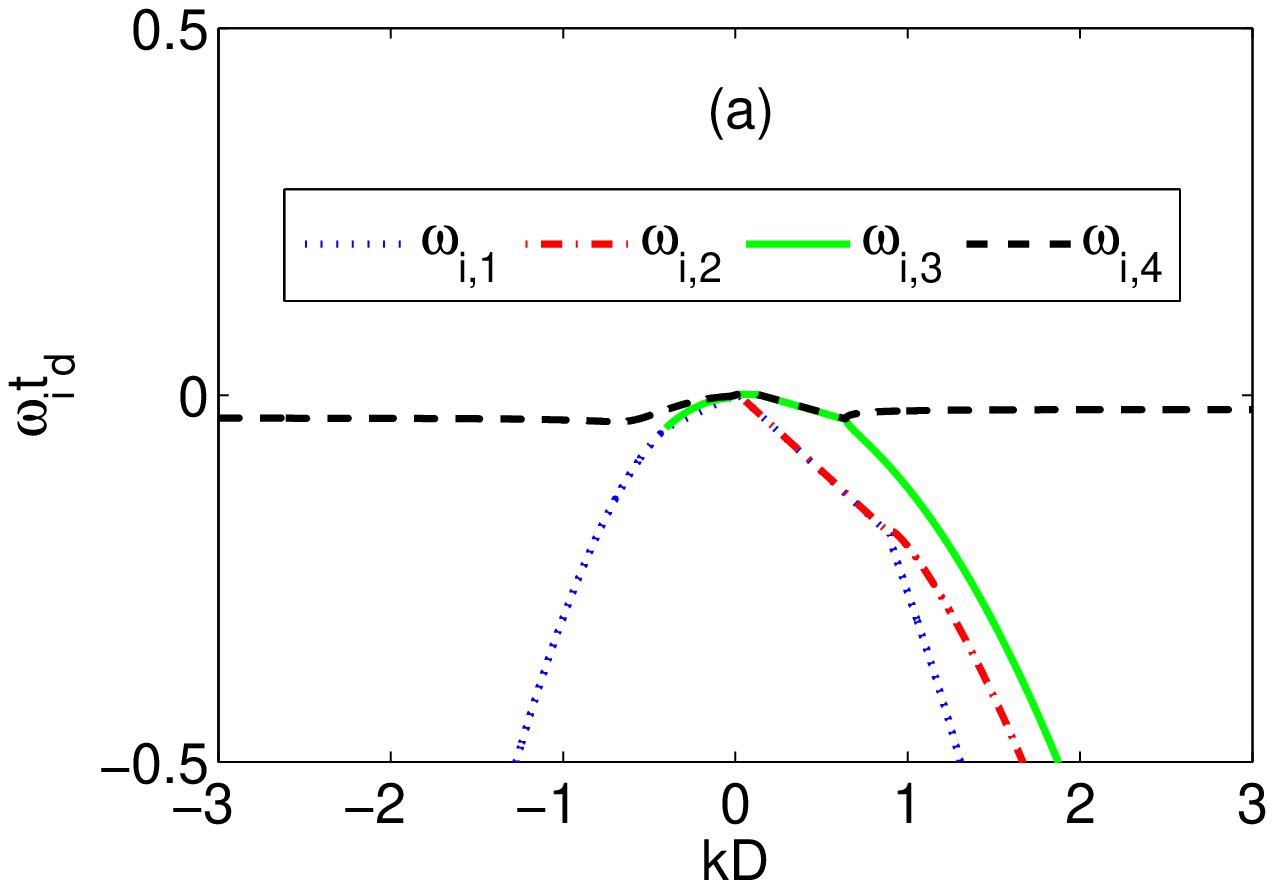}
    \end{center}
   \end{minipage} \hfill
   \begin{minipage}[c]{\textwidth}
    \begin{center}
      \includegraphics[scale=0.60]{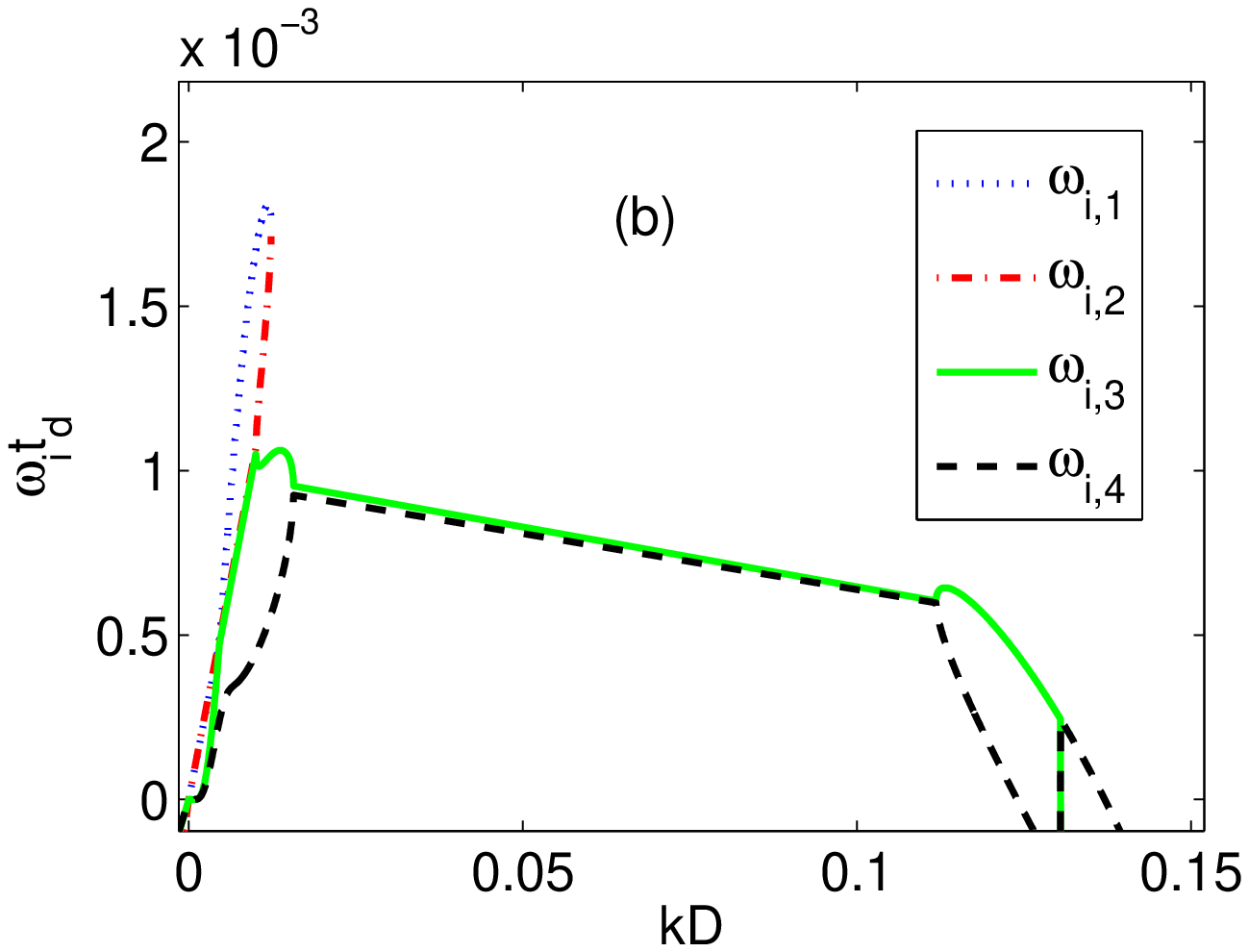}
    \end{center}
   \end{minipage}
\caption{Growth rate $\omega_i$ normalized by the characteristic time $t_d$ as a function of the wavenumber $k$ normalized by the tube diameter $D$. $\omega_{i,1}$ to $\omega_{i,4}$ correspond to the real roots of Eq. \ref{eq:omega_i}}
\label{fig:result1}
\end{figure}

Figure \ref{fig:result1} shows the growth rate $\omega_i$ normalized by the characteristic time $t_d$ as a function of the wavenumber $k$ normalized by the tube diameter $D$. The employed characteristic time is $t_d=18\mu_a/(gd\rho_s)$ and corresponds to the time that a single grain takes to fall the distance equal to its diameter in air. Figure \ref{fig:result1}a, for the broad range of wavenumbers, shows that small wavelengths are stable. This corroborates the long wave assumption in section \ref{section:analysis}.

Figure \ref{fig:result1}b shows the $0\lesssim kD\lesssim 0.15$ region. The figure shows the existence of a most unstable mode and a cut-off wave-number. Only waves within the $0\lesssim kD\lesssim 0.15$ range are unstable and can give rise to plugs. This corresponds to wavelengths in the $0\lesssim \lambda /D\lesssim 40$ range. Given the model uncertainties, the unstable range is considered for the possible appearance of wavelengths instead of only the most unstable mode.

The experimental images were post-processed and the plugs wavelengths were determined. Table \ref{tabela1} summarizes the experimental results. It presents, for each test run, the employed diameter range $d$, the room relative humidity $RH$, the room temperature $T$, the granular mass flow rate $W$, the camera frequency $f$, the mean wavelength $\lambda$ of the granular plugs, the normalized mean wavelength $\lambda/D$ and the normalized standard deviation $\sigma_{\lambda}/D$.

\begin{table}[htbp]
\begin{center}
\begin{tabular}{|c|c|c|c|c|c|c|c|c|c|}
	\hline
	run & $d$ & $RH$ & $T$ & $W$ & $f$ & $\lambda$ & $\lambda/D$ & $\sigma_{\lambda}/D$\\
	$\cdots$ & $\mu m$ & $\%$ & $^oC$ & $g/s$ & $Hz$ & $mm$ & $\cdots$ & $\cdots$ \\
	\hline
	1 & $212-300$ & $42$ & $29$ & $0.52$ & $250$ & $21.4$ & $7.1$ & $1.7$ \\
	\hline
	2 & $212-300$ & $43$ & $26$ & $0.52$ & $250$ & $26.6$ & $8.9$ & $1.3$\\
	\hline
	3 & $212-300$ & $44$ & $26$ & $0.55$ & $250$ & $13.6$ & $4.5$ & $0.8$ \\
	\hline
	4 & $212-300$ & $41$ & $29$ & $0.67$ & $250$ & $9.2$ & $3.1$ & $0.8$ \\
	\hline
	5 & $212-300$ & $34$ & $26$ & $0.39$ & $300$ & $24.9$ & $8.3$ & $5.2$ \\
	\hline
	6 & $106-212$ & $41$ & $25$ & $0.60$ & $250$ & $14.6$ & $4.9$ & $1.5$ \\
	\hline
	7 & $106-212$ & $41$ & $25$ & $0.60$ & $250$ & $11.9$ & $4.0$ & $1.2$ \\
	\hline
	8 & $106-212$ & $41$ & $25$ & $0.80$ & $250$ & $32.7$ & $10.9$ & $1.5$ \\
	\hline
	9 & $106-212$ & $43$ & $25$ & $0.80$ & $250$ & $25.6$ & $8.5$ & $1.5$ \\
	\hline
	10 & $106-212$ & $41$ & $24$ & $0.74$ & $250$ & $19.5$ & $6.5$ & $2.0$ \\
	\hline
	
\end{tabular}
\caption{Grains diameter $d$, room relative humidity $RH$, room temperature $T$, granular mass flow rate $W$, camera frequency $f$, mean wavelength $\lambda$ of the granular plugs, normalized mean wavelength $\lambda/D$ and normalized standard deviation $\sigma_{\lambda}/D$ for each test run.}
\label{tabela1}
\end{center}
\end{table}

The experimental data shows that the plug wavelength is in the range $3\, <\,\lambda /D\, <\, 11$, which is in perfect agreement with the proposed model. However, as only one tube diameter and one grain type were employed, we compare next the present results with previous published experimental results. 

In a series of papers, Raafat et al. (1996) \cite{Raafat}, Aider et al. (1999) \cite{Aider}, and Bertho et al.(2002) \cite{Bertho_1} presented experiments of granular flows in a tube. In particular, concerning the characteristics of density waves, Raafat et al. (1996) \cite{Raafat} reported that the size of plugs was $\lambda /D \approx 10$ and that it was roughly independent of the flow rate. Bertho et al. (2002) \cite{Bertho_1} also reported that the size of plugs was $\lambda /D \approx 10$. In addition, they showed that the wavelength of bubbles is $\lambda_{bubble} /D \approx 10$. These measurements are in agreement with the wavelengths predicted by the proposed model.

The final observation concerns the lowest plug. Bertho et al. (2002) \cite{Bertho_1} reported that at the lower portion of the tube (tube exit) a different plug forms. The length of this plug varies with the flow rate. For $W$ from $1,75\,g/s$ to $3,9\,g/s$, they found that the length of the bottom plug varies from $\lambda /D \approx 200$ to $\lambda /D \approx 30$. This plug is subject to exit boundary conditions, therefore, its wavelength is not correctly predicted by the present analysis.

\section{Conclusions}
\label{section:conclusions}

This paper discussed the density waves that appear when fine grains fall through a tube. The main objective of the paper was the prediction of the wavelength of the high-density regions (plugs) of granular flow. The paper presented a linear stability analysis based on the one-dimensional model proposed by Bertho et al. (2003) \cite{Bertho_2}. The Bertho et al. model was modified by including capillary effects and the closure equations for granular stresses. The fourth-order polynomial equation for the growth rate was numerically solved and the results showed the growth of long waves. The model length scale and the experimental data were in good agreement.

\begin{acknowledgements}
Erick de Moraes Franklin is grateful to FAPESP (grant n. 2012/19562-6) and to FAEPEX/UNICAMP (conv. 519.292, projects AP0008/2013 and 0201/14). Carlos Alvarez Zambrano is grateful to SENESCYT. The authors thank Rodolfo M. Tomazela for the help with the experimental device.
\end{acknowledgements}


\bibliography{referencias}
\bibliographystyle{spphys}

\end{document}